\documentclass{article}
\usepackage{graphicx}
\usepackage{amsmath}
\usepackage{amsfonts}
\usepackage{amssymb}
\newtheorem{theorem}{Theorem}

\newtheorem{proposition}[theorem]{Proposition}

\newenvironment{proof}[1][Proof]{\textbf{#1.} }{\ \rule{0.5em}{0.5em}}

\begin{document}

\title{The Fuzzy Analog of Chiral Diffeomorphisms in higher dimensional Quantum Field Theories}
\author{Lucio Fassarella and Bert Schroer\thanks{Prof. emeritus of the FU-Berlin}\\CBPF, Rua Dr. Xavier Sigaud, 150, Rio de Janeiro - RJ, Brazil\\email Fassarel@cbpf.br,\thinspace\thinspace\ Schroer@cbpf.br}
\date{June 2001}
\maketitle
\begin{abstract}
The well-known fact that classical automorphisms of (compactified) Minkowski
spacetime (Poincar\'{e} or conformal trandsformations) also allow a natural
derivation/interpretation in the modular setting (in the operator-algebraic
sense of Tomita and Takesaki) of the algebraic formulation of QFT has an
interesting nontrivial chiral generalization to the diffeomorphisms of the
circle. Combined with recent ideas on algebraic (d-1)-dimensional lightfront
holography, these diffeomorphisms turn out to be images of ``fuzzy'' acting
groups in the original d-dimensional (massive) QFT. These actions do not
require any spacetime noncommutativity and are in complete harmony with
causality and localization principles.

Their use tightens the relation with kinematic chiral structures on the causal
horizon and makes recent attempts to explain the required universal structure
of a possible future quantum Bekenstein law in terms of Virasoro algebra
structures more palatable.
\end{abstract}

\section{Introduction}

Chiral quantum field theory owes most of its analytic accessibility to the
presence of a (centrally extended) covering of the diffeomorphism group of the
circle. It is well-known that the peculiarity of two-dimensional conformal
automorphisms of Minkowski spacetime shows up in the classical calculation of
conformal symmetries which apart from an x-dependent factor leave the
Minkowski-metric invariant.

These classical arguments have however no direct bearing on symmetries of a
purely quantum origin, i.e. on symmetries which do not enter local quantum
physics via quantization as Poincar\'{e}- or conformal- symmetry, but rather
require the noncommutative\footnote{We use this terminology here in the
old-fashioned sense of the operator theory describing real time quantum
physics being nonabelian. The new sense of a conjectured spacetime
noncommutativity would be verbally compatible with the old one if one adds the
word ``Euclidean'' ore ``spacetime'' before noncommutative.} quantum physical
structure for their very existence. In the standard setting of quantum field
theory such symmetries would correspond to ``fuzzy'' transformations in test
function space $f\rightarrow f^{t}$ which are not representable by
diffeomorphism $T$%
\begin{align}
&  A(f)=\int f(x)A(x)d^{d}x\rightarrow A^{t}(f)\equiv A(f^{t})\\
&  \nexists\,\,T\,\,with\,\,f^{t}(x)=f(Tx)\nonumber
\end{align}
The best mathematical way to describe these symmetries is however via the use
of local operator algebras in the setting of algebraic QFT. Of course in order
to describe a physical symmetry the transformation must satisfy certain
causality restriction and be unitarily implementable with respect to a
reference state (usually the vacuum) in the sense of a Wigner symmetry.

There is a difference between the present ''fuzziness'' and that occurring in
the presently popular noncommutative spacetime theories (related to
noncommutative geometry)\footnote{The motivation for both the algebraic
formulation and the noncommutative spacetime modification of QFT is similar:
the avoidance of the ultraviolet problems. Whereas the algebraic approach
tries to achieve this by a reformulation which bypasses the singular
field-coordinatizations of the standard method, the noncommutative formalism
amounts to a step into the potentially interesting conceptual ``blue yonder''
\cite{Dop}.}. Here it may be helpful to add the following remarks The
motivation for both the algebraic formulation and the noncommutative spacetime
modification of QFT is similar: the avoidance of the ultraviolet problems.
Whereas the algebraic approach tries to achieve this by a reformulation which
bypasses the singular field-coordinatizations of the standard method, the
noncommutative formalism amounts to a step into the potentially interesting
conceptual ``blue yonder'' \cite{Dop}.

In the present case the pointlike geometric action of the Poincar\'{e} or
conformal group is maintained since the indexing by spacetime regions in a net
of noncommutative algebras remains perfectly classical, although this
algebraic setting reveals an unexpected interplay between this classical
indexing and the very noncommutative aspects of the algebraic modular aspects.
The fuzziness refers to the modular analogs of spacetime transformations which
cannot be encoded into diffeomorphisms of spacetime (but which may have some
geometric interpretation in terms of the infinite dimensional geometry of
certain real subspaces of a complex Hilbert space (in the absence of
interactions such investigations have been carried out in \cite{JMP}%
\cite{BGL}). The physical significance of pure quantum symmetries of non
Noetherian origin is presently under investigation but will not be discussed here.

Instead of starting an axiomatic discussion of which fuzzy transformations are
physically admissible, we find it more natural to go directly into medias res
and present the only known mechanism which produces such fuzzy
transformations: the modular localization theory of local quantum physics. It
attaches a unitary one-parametric symmetry $\Delta_{\mathcal{O},\Omega}^{it}$
to each standard pair ($\mathcal{A}(\mathcal{O}),\Omega)$ of an operator
algebra which in the standard formulation of QFT would be affiliated with
generating smeared fields (with testing functions with support in the causally
closed region $\mathcal{O})$ together with a cyclic and separating vector
$\Omega.\cite{Borchers}$ This transformation does respect the causal horizon
of $\mathcal{O}$ and maintains causal disjointness by not mixing $\mathcal{O}$
with $\mathcal{O}^{\prime},$ but apart from this compatibility with causality
with respect to a prescribed region it acts in the above described fuzzy
manner. For certain special localization region of algebras, namely in case of
massive theories for a noncompact wedge region $W$ instead of a compact double
cone (and for massless theories which lead to conformal invariance even in
case of a double cone $\mathcal{O)}$, the modular transformation
$\Delta_{\mathcal{O},\Omega}^{it}$ with respect to the vacuum becomes a local
Lorentz boost (or a special conformal transformation). In that case one learns
that in Poincar\'{e} (or conformal) transformations well-known from the
quantization of Noether's theorem can also be generated from a finite number
of modular symmetries; in fact for $d>1+1$ they are the only geometric modular
symmetries \cite{Kaehler}. In fact the inverse mapping of the holographic
isomorphism in the illustration (attached after references) would
automatically carry chiral symmetries which are not pure diffeomorphisms but
also involve charge measuring currents (as e.g. chiral loop groups) into fuzzy
symmetries of the massive d=1+1 model. Knowing the Doplicher-Roberts theory
\cite{DR}, which explains the standard internal group symmetry of particle
physics in terms of (apparently quite unrelated with group theory) causality
and spectral properties of local observables, the idea that even inner
symmetries are rooted in their representation theory via localizable states is
less outrageous as it appears at first sight. In this paper we will however
limit our attention to pure diffeomorphisms.

\textit{In this work we would like to argue that the infinite dimensional
modular group }$G$\textit{\ generated by the fuzzy }$\Delta_{\mathcal{O}%
,\Omega}^{it}$\textit{\ in the common Hilbert space }$H$\textit{\ for all
double cones }$O$\textit{\ (}and possibly\textit{\ }$\Omega$ standing for more
states than just the vacuum\textit{) is the physically correct substitute for
a nonexistent infinite dimensional geometrically acting diffeomorphism group.
}In order for such a claim to be make sense we must first show that the
well-known chiral diffeomorphism group (whose infinitesimal generators obey
the Virasoro algebra) have a modular origin.

For this reason the algebraic quantum field theory setting of this paper is
crucial and we hope to convince the reader that this conceptually novel way of
looking at QFT is more than just a pastime of some specialists who are less
than satisfied with the standard way of doing QFT.

The transition from standard QFT to AQFT is similar to that of geometry in
coordinates to the modern intrinsic coordinate-free formulation with the
pointlike fields corresponding to the use of coordinates. But contrary to
coordinate-free geometry, which still uses coordinate patches in the
formulation of manifolds and allows to derive all results in principle by
(often clumsy) calculation in coordinates, it is less clear that every
algebraically defined QFT allows a complete description in terms of pointlike
field generators. For chiral theories and presumably also for higher
dimensional conformal field theories the algebraic and pointlike formulations
are known to be equivalent \cite{Joerss}. The importance of the algebraic
method would not be diminished if pointlike field descriptions exist on both
sides of the holography, since the field coordinatization of the original
theory turns out to be radically different from that of its holographic image.

We illustrate these points in an appendix by d=1+1 examples of massive models.
It should be already clear from our use of terminology that the holography in
this paper is primarily treated as quantum phenomenon in Minkowski space. In a
way it supersedes lightray/lightfront quantization and the $p\rightarrow
\infty$ frame method in a way which will be touched on later (for a more
detailed discussion see \cite{holo}).

The enhancement of quantum symmetries to (generally ``fuzzily acting'')
infinite symmetry groups is also very interesting from a mathematical
viewpoint. The quest of the mathematicians as to what constitutes the most
useful generalization of finite dimensional Lie group theory to infinite
dimensions has found a successful answer in the theory of loop groups (more
generally Kac-Moody algebras) and associated diffeomorphism groups of the
circle (Virasoro algebras). However the physical use of these mathematical
findings is restricted to low dimensional field theory where they make a
perfect match with the causality aspects of current or energy-momentum
operators \cite{Kac}. Therefore it is natural to ask whether the causality and
localization principles, together with some other mathematical vehicle which
does not confine them to low spacetime dimensions, could not produce a more
general framework of symmetry- and group generalizations. Our tentative answer
is that, thanks to the existence of the Tomita-Takesaki modular theory
\cite{Tomita}, this seems to be indeed possible.

Physicists are used to distinguish \textit{inner symmetries} related to
charges of particles from \textit{spacetime symmetries }(often called
outer)\footnote{Outside of the standard approach to QFT one canot expect a
sharp separation between these two symmetries.}. In the first case the
generalization has led to the relevance of braid groups, knot theory,
superselection sector theory, endomorphism- and subfactor- theory. An
important mathematical step was the study of Jones inclusions and the ensuing
basic construction \cite{Jones} which is intimately related to modular theory.
For the generalized spacetime symmetries, the important step was to first
unravel the modular origin of the geometric symmetries (Poincar\'{e},
conformal) and then to understand in which sense the infinitely many fuzzy
modular symmetries can be seen as an answer to the above question. In this
study the role of the Jones basic construction and subfactors (which in
physics is related to generalized inner symmetries) is replaced by the notion
of \textit{modular inclusions \cite{Wies}}.

Our present proposal is belonging to this second kind of spacetime symmetry
generalization which turns out to be closely related to a conceptually concise
formulation of holography. In fact it replaces the old light cone quantization
which was extremely formal, and had in addition to those problems which it
shares with the canonical quantization (restriction to canonical operator
dimensions which creates a clash with genuine interactions) additional
causality problems of its own in that it remained obscure in what sense the
original local QFT can be reconstructed from the data of light cone
quantization (or the p$\rightarrow\infty$ frame method).

In the next section we will confirm the validity of this modular
interpretation of chiral diffeomorphisms in the case of the abelian current
model. In particular we will construct a vector $\Phi$ in the Hilbert space
which, together with chiral algebra $\mathcal{A}(0,1)$ localized in the
interval $\left[  0,1\right]  ,$ generates a diffeomorphism subgroup
$D_{2}(\lambda)$ with 4 fixed points which leaves $\Phi$ invariant and is in
fact the Tomita-Takesaki modular group of ($\mathcal{A}(0,1),\Phi$). This
should be compared with the well-known fact \cite{Borchers} that the standard
Moebius dilation $D(\lambda)$ with the two fixed points $0,\infty$ is the
modular group of the standard pair $(\mathcal{A}(0,\infty),\Omega)$ where from
now on $\Omega$ denotes the vacuum state vector. The generator of this
$D_{2}(\lambda)$ dilation together with those of the modular generated
M\"{o}bius group generate the full system of $L_{n}$-generators $n\in
\mathbb{Z}.$

The third section generalizes this observation to arbitrary chiral theories.

In the fourth section we show that among the infinitely many fuzzy
transformations in higher dimensional QFT which are of modular origin, there
is an important subset which agrees with the above diffeomorphism but only if
we ``holographically reprocess'' the original theory.

The necessary lightfront holography is more radical than the better known (but
very special) algebraic AdS-CQFT holography \cite{Rehren} as well as with the
recently proposed concept of ``transplantation'' \cite{BMS}, but its intuitive
basis is still similar to the original pre-Ads work on holography by 't Hooft
which demands such a radical spacetime ``scrambling up'' of degrees of freedom
while maintaining their number size. In particular there is no lower
dimensional asymptotic region for which the pointlike correlation functions of
the original theory coalesce with those of the holographic image \cite{MW}.
Whereas in the AdS-CQFT isomorphism, as a result of its shared high symmetry,
one needs the algebraic reprocessing (which avoids field coordinatizations)
mainly in the \textit{unique structural determination of the higher
dimensional AdS}$_{d+1}$\textit{\ theory in terms of its lower dimensional
CQFT}$_{d}$\textit{\ holographic image \cite{Rehren}}, in more general and
less symmetric cases, in particular when the holographic image is localized on
the causal horizon of a higher dimensional region, the algebraic method
becomes \textit{indispensable for both directions} of the holographic
processing \cite{AdS} \cite{holo}.

\section{A mathematically controllable illustration}

Let us first investigate the modular origin of the diffeomorphism group of the
circle for the Weyl algebra. Following \cite{S-W} we represent these
diffeomorphism in terms of the standard dilation $D(\lambda)$ sandwiched
between stretching and its inverse compression transformation%

\begin{align}
&  \xi:\mathbb{(}0,1\mathbb{)}\rightarrow(0,\infty)\;,\;\xi(x):=\frac
{2x}{1-x^{2}}\,\,\\
&  with\,\,inverse\,\,\xi_{(0,1)}^{-1}(x)=\frac{\sqrt{x^{2}+1}-1}{x}=-\frac
{1}{x}+\sqrt{1+\frac{1}{x^{2}}}\;\nonumber\\
&  D_{2}(\lambda)x=\left(  \xi^{-1}D(\lambda)\xi\right)  (x)=-\frac{1-x^{2}%
}{2\lambda x}+\sqrt{1+\frac{\left(  1-x^{2}\right)  ^{2}}{4\lambda^{2}x^{2}}%
}\nonumber\\
&  D_{2}(\lambda):(0,1)\rightarrow(0,1),\,\,extendible\,\,to\text{
}diff.\text{ }of\text{\thinspace\thinspace}R_{comp}\nonumber
\end{align}
The $\xi$-transformations are the Cartesian versions of the circular
transformations $z\rightarrow z^{2},z\rightarrow\sqrt{z},$ which although not
diffeomorphism of the circle themselves, do ``lift'' the ordinary M\"{o}bius
transformations $Moeb$ with two fixed points to its quasisymmetric (in the
mathematical sense of complex function theory) analog $Moeb_{2}$ with the
$D_{2}(\lambda)$ the analog of $D(\lambda)$ having four fixed points instead
of two. In the Cartesian description the $\lambda\rightarrow\infty$ limits are
somewhat easier to handle.

It is not difficult to find a state of the current algebra which is invariant
under $Moeb_{2},$ one only has to modify the vacuum two point function of the
current by a factor\footnote{The following calculations are identical to those
in section 3 of \cite{S-W}. The use of these calculations for the
determination of the \textit{modular theory of multi-interval}s is however
incorrect since the state $\left\langle \cdot\right\rangle _{2}$ ceases to be
faithful on large algebras containing oppositely localized intervals. This is
the reason for the present restriction of test functions.}
\begin{equation}
\left\langle j(x)j(y)\right\rangle _{2}=\left\langle j(x)j(y)\right\rangle
_{0}\frac{\left(  1+x^{2}\right)  (1+y^{2})}{\left(  1+xy\right)  ^{2}}
\label{state}%
\end{equation}
and one easily checks that thanks to this additional factor the new
expectation value is invariant under $Moeb_{2},$ in particular under
$D_{2}(\lambda).$ But since the state is not faithful on algebras whose
localization includes opposite intervals, we have to restrict the algebra
generated by currents localized in the interval ($0,1$)$.$ Whereas the
transformation $D_{2}(\lambda)$ is globally well-defined, the state
$\left\langle j(x)j(y)\right\rangle _{2}$ is only faithful and normal on
subalgebras which are localized between two fixed points of $D_{2}(\lambda)$.
In order to have a formulation in terms of bounded operators one may pass to
the Weyl algebra in which case the above formula defines a state $\left\langle
\cdot\right\rangle _{2}$ on the von Neumann algebra
\begin{align}
\mathcal{A}(0,1)  &  \equiv alg\left\{  W(f)|\,suppf\subset\left[  0,1\right]
\right\} \label{Weyl}\\
W(f)  &  =e^{ij(f)},\,j(f)=\int f(x)j(x)dx\nonumber\\
\left\langle W(f)\right\rangle _{2}  &  =e^{-\frac{1}{2}\left\langle
j(f)j(f)\right\rangle _{2}}\nonumber
\end{align}
The invariance of this algebra and the state under $D_{2}(\lambda
)$-transformations is expressed by the identity
\begin{align}
&  \int\frac{f_{\lambda,2}(x)g_{\lambda,2}(y)}{\left(  x-y+i\varepsilon
\right)  ^{2}}\frac{\left(  1+x^{2}\right)  (1+y^{2})}{\left(  1+xy\right)
^{2}}dxdy=\\
&  \int\frac{f(x)g(y)}{\left(  x-y+i\varepsilon\right)  ^{2}}\frac{\left(
1+x^{2}\right)  (1+y^{2})}{\left(  1+xy\right)  ^{2}}dxdy\nonumber\\
&  f_{\lambda,2}(x)\equiv f(D_{2}(\lambda)x)=\left(  D_{2}^{-1}(\lambda)\circ
f\right)  (x),\,suppf\subset\left[  0,1\right] \nonumber
\end{align}
The intermediate steps of the calculation can be found in \cite{S-W} page 146.

For the following it is convenient to rewrite the state in terms of the vacuum
state and the previously introduced map $\xi^{-1}$%
\begin{align}
&  \int\frac{f(x)g(y)}{\left(  x-y+i\varepsilon\right)  ^{2}}\frac{\left(
1+x^{2}\right)  (1+y^{2})}{\left(  1+xy\right)  ^{2}}dxdy=\label{form}\\
&  \int\frac{\left(  \xi\circ f\right)  (x)(\xi\circ g)(y)}{\left(
x-y+i\varepsilon\right)  ^{2}}dxdy,\,\,\,\left(  \xi\circ f\right)  (x)\equiv
f(\xi^{-1}(x))\nonumber\\
&  or\,\,\,\left\langle j(f)j(g)\right\rangle _{2}=\left\langle j(\xi\circ
f)j(\xi\circ g)\right\rangle \nonumber
\end{align}
In this way of writing the KMS property of the new state with respect to the
action of $D_{2}(\lambda)$ on $\mathcal{A}(0,1)$ becomes obvious and follows
from the identity \cite{S-W}
\begin{align}
\left\langle j(D_{2}^{-1}(\lambda)\circ f)j(g)\right\rangle _{2}  &
=\left\langle j(D^{-1}(\lambda)\circ\xi\circ f)j(\xi\circ g)\right\rangle \\
suppf  &  \subset\left[  0,1\right]  ,\,\,supp\xi\circ f\subset\left[
0,\infty\right] \nonumber
\end{align}
together with the vacuum KMS property of the right hand side \cite{Yng}. The
right hand side has the strip analytic properties in $t\rightarrow t+i$ where
$\lambda=e^{2\pi t}$ and the values on the upper rim of the strip are related
to those of the lower by the KMS relation. With the KMS prerequisite for the
action of $D_{2}(\lambda)$ on $\mathcal{A}(0,1)$ in the state $\omega
_{2}\equiv\left\langle \cdot\right\rangle _{2}$ being fulfilled, one only has
to find a globally $D_{2}(\lambda)$-invariant vector $\Phi$ in $H$ which
implements $\omega_{2}$%
\begin{align}
\left\langle A\right\rangle _{2}  &  =\left\langle \Phi\left|  A\right|
\Phi\right\rangle \\
A  &  \in\mathcal{A}(0,1)\nonumber
\end{align}
For this purpose we use the fact that the faithful state $\omega_{2}$ on the
algebra $\mathcal{A}(0,1)$ has a unique vector implementation in the natural
cone $\mathcal{P}_{\Omega}$ of the standard pair $(\mathcal{A}(0,1),\Omega)$
\cite{Haag} with a vector $\Phi$ which is automatically separating with
respect to $\mathcal{A}(0,1)$ as a result of the faithfulness of $\omega_{2}$
on $\mathcal{A}(0,1).$ This globally defined vector inherits the invariance
under the $D_{2}(\lambda)$ action from $\omega_{2}$ and hence $D_{2}%
(\lambda=2\pi t)$ is the unitary modular operator $\Delta_{(\mathcal{A}%
(0,1),\Phi)}^{it}$ of $(\mathcal{A}(0,1),\Phi).$ The expectation values of
operators localized outside the interval $(0,1)$ are only known after an
explicit calculation of $\Phi\footnote{Most of the calculations in modular
theory are to prove existence and structural properties whereas explicit
constructions are usually ver hard.}.$ They are certainly not given by the
extension of the state $\omega_2$ (\ref{Weyl}) outside $\mathcal{A}(0,1).$

Since the automorphism induced by the Ad action of $D_{2}$ on the test
functions is identical to that of $\frac{1}{2}(L_{2}-L_{-2})$ in standard
notation and the commutator with $L_{0}$ is proportional to the sum
$L_{2}+L_{-2}$ and hence one obtains all the generators of $Moeb$ and
$Moeb_{2}$ through commutators. It has been shown that $Moeb$ has a modular
origin \cite{Wies}\cite{Longo}. From this, together with the fact that all the
$L_{\pm n}$ for $n>2$ are obtainable iteratively from $L_{\pm1}$ and $L_{\pm
2},$ follows our claim that the diffeomorphism group of the chiral
current-Weyl algebra can be obtained by modular methods. This is very
interesting since modular methods to not suffer from restrictions in spacetime
dimensions as those methods which use the structure of the chiral energy
momentum tensor or the chiral current algebra.

In order to elevate this model observation into a new tool of local quantum
physics, we still have to overcome to hurdles. First we should generalize this
observation to arbitrary chiral theories and second we should understand the
higher dimensional analog.

\section{Extension to general chiral theories}

We now apply similar considerations to general chiral theories. For such
models it has been shown that there is not much difference between the
pointlike and the algebraic viewpoint; algebraic nets always may be considered
as being generated by pointlike fields \cite{Joerss}. For simplicity of
presentation we will use the better known pointlike formulation.

In order to show the modular origin of the $D_{2}(\lambda)$-symmetry we study
(for simplicity) the sequence of Wightman functions of the $D_{2}(\lambda)$
transforms of a (primary) field $D_{2}(\lambda)$ $\phi$ with dimension $d$ and
show that the limit $\lambda\rightarrow\infty$ exists and is $D_{2}(\lambda)$
invariant. This can be used as an alternative method to show the existence of
unitary implementation of the $D_{2}(\lambda)$-automorphism.

The M\"{o}bius-invariance of the vacuum can be expressed in terms of
observable\footnote{For fields with anomalous dimensions which live on the
covering of the conformal compactification this simple invariance relation is
restricted to those conformal transformations which leave infinity invariant
\cite{Sch}.} correlation functions
\begin{equation}
\left\langle \phi(x_{1})...\phi(x_{m})\right\rangle _{0}=\left[  g^{\prime
}(x_{1})...g^{\prime}(x_{m})\right]  ^{d}\left\langle \phi(g(x_{1}%
))...\phi(g\left(  x_{m}\right)  )\right\rangle _{0},\,\,g\in Moeb \label{qpf}%
\end{equation}

Using the noninvariance of the vacuum under $D_{2}(\lambda)$ we define the
following $\lambda$-dependent sequence of correlation functions
\begin{align*}
\mathcal{\varphi}_{\lambda}(x_{1},...,x_{m})  &  :=\left[  D_{2}%
(\lambda)^{\prime}(x_{1})\right]  ^{d}...\left[  D_{2}(\lambda)^{\prime}%
(x_{m})\right]  ^{d}\times\\
&  \times\left\langle \phi(D_{2}(\lambda)\left(  x_{1}\right)  )...\phi\left(
D_{2}(\lambda)(x_{m})\right)  \right\rangle _{0}%
\end{align*}

The following formula for $D_{2}(\lambda)$ is suitable for taking the limit
$\lambda\rightarrow\infty$:
\begin{equation}
D_{2}(\lambda)x=1-\frac{1/\xi(x)}{\lambda}+\eta(\lambda,x)\;,\;\forall
x\in(0,1) \label{D2expand}%
\end{equation}
where $\eta$ is a smooth function of $x$ for fixed $\lambda$ and is of order
$\lambda^{-2}$ for fixed $x$. In particular:
\begin{equation}
\lim_{\lambda\rightarrow\infty}\lambda\eta(\lambda,x)=0\;,\;\forall x\in(0,1)
\label{eta}%
\end{equation}

We also have
\begin{equation}
D_{2}(\lambda)^{\prime}x=\frac{\xi^{\prime}(x)/\xi^{2}(x)}{\lambda}%
+\lambda\eta^{\prime}(\lambda,x)\xi^{\prime}(x) \label{D2'expand}%
\end{equation}
and
\begin{equation}
\lim_{\lambda\rightarrow\infty}\lambda^{2}\eta^{\prime}(\lambda,x)=0
\label{eta'}%
\end{equation}

We now can rewrite $\mathcal{\varphi}_{\lambda}$:

%
%
%
%
%
\begin{align}
&  \mathcal{\varphi}_{\lambda}(x_{1},...,x_{m}):=\left[  \frac{\xi^{\prime
}(x_{1})/\xi^{2}(x_{1})}{\lambda}+\lambda\eta^{\prime}(\lambda,x_{1}%
)\xi^{\prime}(x_{1})\right]  ^{d}...\times\\
&  \hspace*{2.3cm}\times\left\langle \phi\left(  1-\frac{1/\xi(x_{1})}%
{\lambda}+\eta(\lambda,x_{1})\right)  ...\right\rangle _{0}\nonumber
\end{align}
\newline 

By using translational invariance of vacuum we obtain
\begin{align}
\mathcal{\varphi}_{\lambda}(x_{1},...,x_{m})  &  =\left[  \frac{\xi^{\prime
}(x_{1})/\xi^{2}(x_{1})}{\lambda}+\lambda\eta^{\prime}(\lambda,x_{1}%
)\xi^{\prime}(x_{1})\right]  ^{d}...\times\\
&  \times\left\langle \phi\left(  -\frac{1/\xi(x_{1})}{\lambda}+\eta
(\lambda,x_{1})\right)  ...\right\rangle _{0}\nonumber\\
&  =\left(  \frac{1}{\lambda}\right)  ^{md}\left[  1\xi^{\prime}(x_{1}%
)/\xi^{2}(x_{1})+\lambda^{2}\eta^{\prime}(\lambda,x_{1})\xi^{\prime}%
(x_{1})\right]  ^{d}...\times\nonumber\\
&  \times\left\langle \phi\left(  \frac{1}{\lambda}\left[  -1/\xi
(x_{1})+\lambda\eta(\lambda,x_{1})\right]  \right)  ...\right\rangle
_{0}\nonumber
\end{align}

Using scaling invariance of vacuum, we get
%
%
%
%
%
\begin{align}
&  \mathcal{\varphi}_{\lambda}(x_{1},...,x_{m})=\left[  \xi^{\prime}%
(x_{1})/\xi^{2}(x_{1})+\lambda^{2}\eta^{\prime}(\lambda,x_{1})\xi^{\prime
}(x_{1})\right]  ^{d}...\times\\
&  \hspace*{2.3cm}\times\left\langle \phi\left[  -1/\xi_{\tilde{n}}%
(x_{1})+\lambda\eta(\lambda,x_{1})\right]  ...\right\rangle _{0}\nonumber
\end{align}

By taking the limit $\lambda\rightarrow\infty$,
%
%
%
%
%
\begin{align}
&  \lim_{\lambda\rightarrow\infty}\mathcal{\varphi}_{\lambda}(x_{1}%
,...,x_{m})=\left[  \xi^{\prime}(x_{1})/\xi^{2}(x_{1})\right]  ^{d}...\left[
\xi^{\prime}(x_{m})/\xi^{2}(x_{m})\right]  ^{d}\times\\
&  \hspace*{2.3cm}\times\left\langle \phi\left[  \left(  -1/\xi(x_{1}\right)
)...\phi\left(  -1/\xi(x_{m}\right)  \right]  \right\rangle _{0}\nonumber
\end{align}

Again, scaling invariance of vacuum implies
\begin{align}
\lim_{\lambda\rightarrow\infty}\mathcal{\varphi}_{\lambda}(x_{1},...,x_{m})
&  =\left[  \left.  \left(  \frac{-1}{\xi}\right)  ^{\prime}\right|
_{(x_{1})}...\left.  \left(  \frac{-1}{\xi}\right)  ^{\prime}\right|
_{(x_{m})}\right]  ^{d}\times\nonumber\\
&  \times\left\langle \phi_{\alpha_{1}}\left(  -1/\xi(x_{1}\right)
)...\phi_{\alpha_{m}}\left(  -1/\xi(x_{m}\right)  )\right\rangle _{0}%
\end{align}

M\"{o}bius invariance of vacuum implies ($x\rightarrow-1/x$ belongs to the
M\"{o}bius group)
%
%
%
%
%
\begin{align}
&  \lim_{\lambda\rightarrow\infty}\mathcal{\varphi}_{\lambda}(x_{1}%
,...,x_{m})=\left[  \left.  \left(  -1/\xi\right)  ^{\prime}\right|
_{(x_{1})}...\left.  \left(  -1/\xi\right)  ^{\prime}\right|  _{(x_{m}%
)}\right]  ^{d}\xi^{2d}(x_{1})...\xi^{2d}(x_{m})\times\\
&  \hspace*{2.3cm}\times\left\langle \phi\left(  \xi(x_{1}\right)
)...\phi\left(  \xi(x_{m}\right)  )\right\rangle _{0}\nonumber
\end{align}
Simplifying, we get the simple expression, analogous to \ref{qpf},
\begin{equation}
\lim_{\lambda\rightarrow\infty}\mathcal{\varphi}_{\lambda}(x_{1}%
,...,x_{m})=\left[  \xi^{\prime}(x_{1})...\xi^{\prime}(x_{m})\right]
^{d}\left\langle \phi\left(  \xi(x_{1}\right)  )...\phi\left(  \xi
(x_{m}\right)  )\right\rangle _{0}%
\end{equation}

We call $\mathcal{\varphi}$ the state over $\mathcal{A}(0,1)$ defined by the
above correlation functions and designed by $\left\langle ...\right\rangle
_{2}$ for latter convenience.

\begin{proposition}
$\mathcal{\varphi}$ is a positive faithful normal state over $\mathcal{A}%
(0,1)$ which is invariant under $D_{2}(\lambda)$ for all $\lambda>0$.

\begin{proof}
The local normality follows directly from the limit construction via a
sequence of normal states. Positiveness of $\mathcal{\varphi}$ is equivalent
to the proposition \cite[Theorem 3-3]{Str-Wi}: for any sequence of test
functions $\left\{  f_{j}\right\}  $, $f_{j}(x_{1},...,x_{j})$ being
$C^{\infty}$ functions defined for $x_{1},...,x_{j}\in$ $(0,1)$ and with
$f_{j}=0$ except for a finite number of $j$'s, it holds the inequalities
%
%
%
%
%
\begin{align}
&  0\leq\sum_{k,j=0}^{\infty}\int_{0}^{1}...\int_{0}^{1}dx_{1}...dx_{j}%
dy_{1}...dy_{k}\bar{f}_{j}(x_{1},...,x_{j})f_{k}(y_{1},...,y_{k})\times\\
&  \hspace*{2.3cm}\times\left\langle \phi(x_{j})...\phi(x_{1})\phi
(y_{1})...\phi(y_{k})\right\rangle _{2}\nonumber
\end{align}

But, if we define the functions
\begin{multline}
\hat{f}_{j}(x_{1},...,x_{j}):=f_{j}(\xi^{-1}(x_{1}),...,\xi^{-1}(x_{j}%
))\times\left(  \xi^{-1}\right)  ^{\prime}(x_{1})...\left(  \xi^{-1}\right)
^{\prime}(x_{j})\;\\
\text{for}\;x_{1},...,x_{j}\in(0,\infty)\nonumber
\end{multline}
we can write the above sum as
%
%
%
%
%
%
\begin{align}
&  \sum_{k,j=0}^{\infty}\int_{0}^{1}...\int_{0}^{1}dx_{1}...dx_{j}%
dy_{1}...dy_{k}\bar{f}_{j}(x_{1},...,x_{j})f_{k}(y_{1},...,y_{k})\times\\
&  \times\left\langle \phi(x_{j})...\phi(x_{1})\phi(y_{1})...\phi
(y_{k})\right\rangle _{2}=\nonumber\\
&  =\sum_{k,j=0}^{\infty}\int...\int dx_{1}...dx_{j}dy_{1}...dy_{k}\bar{f}%
_{j}(x_{1},...,x_{j})f_{k}(y_{1},...,y_{k})\times\nonumber\\
&  \;\;\;\;\;\;\;\;\;\;\;\;\;\;\times\left[  \xi^{\prime}(x_{1})...\xi
^{\prime}(y_{k})\right]  ^{d}\left\langle \phi(\xi(x_{j}))...\phi(\xi
(x_{1}))\phi(\xi(y_{1}))...\phi((\xi(y_{k}))\right\rangle _{0}\nonumber\\
&  =\sum_{k,j=0}^{\infty}\int...\int dx_{1}...dx_{j}dy_{1}...dy_{k}\bar{f}%
_{j}(\xi^{-1}(x_{1}),...)f_{k}(\xi^{-1}(y_{1}),...)\times\nonumber\\
&  \times\left(  \xi^{-1}\right)  ^{\prime}(x_{1})...\left(  \xi^{-1}\right)
^{\prime}(x_{j})\left(  \xi^{-1}\right)  ^{\prime}(y_{1})...\left(  \xi
^{-1}\right)  ^{\prime}(y_{k})\times\nonumber\\
&  \times\left[  \xi^{\prime}(\xi^{-1}(x_{1}))...\xi^{\prime}(\xi^{-1}%
(y_{k}))\right]  ^{d}\left\langle \phi(x_{j})...\phi(x_{1})\phi(y_{1}%
)...\phi(y_{k})\right\rangle _{0}\nonumber\\
&  =\sum_{k,j=0}^{\infty}\int...\int dx_{1}...dx_{j}dy_{1}...dy_{k}\bar{f}%
_{j}(\xi^{-1}(x_{1}),...)f_{k}(\xi^{-1}(y_{1}),...)\times\nonumber\\
&  \times\left\langle \phi(x_{j})...\phi(x_{1})\phi(y_{1})...\phi
(y_{k})\right\rangle _{0}\nonumber\\
&  =\sum_{k,j=0}^{\infty}\int_{0}^{1}...\int_{0}^{1}dx_{1}...dx_{j}%
dy_{1}...dy_{k}\overline{\hat{f}}_{j}(x_{1},...,x_{j})\hat{f}_{k}%
(y_{1},...,y_{k})\times\nonumber\\
&  \hspace*{2.3cm}\times\left\langle \phi(x_{j})...\phi(x_{1})\phi
(y_{1})...\phi(y_{k})\right\rangle _{0}\nonumber
\end{align}
which turns out to be non-negative as a result of vacuum
positivity.\footnote{The first equality, follows directly from definition of
$\left\langle \;\right\rangle _{\Phi}$; the second equality follows from
change of variables $x\rightarrow\xi^{-1}(x)$; and the thirty equality follows
from definition of the $\hat{f}$'s.}

From a similar expression, we can see $\mathcal{\varphi}$ is faithful on
$\mathcal{A}(0,1)$ since vacuum is faithful on $\mathcal{A}(0,\infty
)$.\thinspace\thinspace First, note that
\[
\xi(D_{2}(\lambda)x)=\lambda\xi(x)
\]
and
\[
\xi^{\prime}(D_{2}(\lambda)x)=\frac{\lambda\xi^{\prime}(x)}{D_{2}%
(\lambda)^{\prime}(x)}%
\]

Then
%
%
%
%
%
\begin{align}
&  \left\langle \phi(D_{2}(\lambda)x_{1})...\phi(D_{2}(\lambda)x_{n}%
)\right\rangle _{2}=\\
&  \left[  \xi^{\prime}\left(  D_{2}(\lambda)x_{1}\right)  ...\xi^{\prime
}\left(  D_{2}(\lambda)x_{n}\right)  \right]  ^{d}\left\langle \phi(\lambda
\xi(x_{1}))...\phi(\lambda\xi(x_{m}))\right\rangle _{0}\nonumber\\
&  =\left[  \frac{\lambda\xi^{\prime}(x_{1})}{D_{2}(\lambda)^{\prime}(x_{1}%
)}...\frac{\lambda\xi^{\prime}(x_{n})}{D_{2}(\lambda)^{\prime}(x_{n})}\right]
^{d}\lambda^{-nd}\left\langle \phi(\xi(x_{1}))...\phi(\xi(x_{n}))\right\rangle
_{0}\nonumber\\
&  =\left[  D_{2}(\lambda)^{\prime}(x_{1})...D_{2}(\lambda)^{\prime}%
(x_{n})\right]  ^{-d}\left\langle \phi(x_{1})...\phi(x_{n})\right\rangle
_{2}\nonumber
\end{align}
\end{proof}
\end{proposition}

From the calculations in the above proof, we see that $D_{2}(\lambda)$ defines
an automorphism on the field $\phi$
\begin{equation}
\left[  \alpha\left(  D_{2}(\lambda)\right)  \phi\right]  (x)=\left[
D_{2}(\lambda)^{\prime}(x)\right]  ^{d}\phi\left(  D_{2}(\lambda)(x)\right)
\end{equation}
and the correlation functions $\left\langle ...\right\rangle _{2}$ are
transformed according to
\begin{equation}
\left\langle \phi(x_{1})...\phi(x_{m})\right\rangle _{2}=\left[  D_{2}%
(\lambda)^{\prime}(x)...D_{2}(\lambda)^{\prime}(x)\right]  ^{d}\left\langle
\phi(D_{2}(\lambda)(x_{1}))...\phi(D_{2}(\lambda)(x_{m}))\right\rangle _{2}%
\end{equation}

Since $\mathcal{\varphi}$ is normal, there exists a unique vector $\left|
\Phi\right\rangle $ in the natural cone $\mathcal{P}_{\Omega}$ of the standard
pair $(\mathcal{A}(0,1),\Omega)$ representing $\mathcal{\varphi}$:
\[
\mathcal{\varphi}(A)=\left\langle \Phi|A|\Phi\right\rangle
\]

Since $\mathcal{\varphi}$ is faithful for $\mathcal{A}(0,1)$, it turns out
that $\Phi$ is cyclic and separating for this algebra. With this vector, we
can extend $\mathcal{\varphi}$ to the entire algebra $\mathcal{A}%
=\mathcal{A}(\mathbb{\dot{R}})=\mathcal{A}(S^{1})$.

To realize that the modular group associated to $\mathcal{\varphi}$ and
$\mathcal{A}(0,1)$ is given by the action of $D_{2}(\lambda)$, we have to
proceed in exactly the same manner as in the special case of current field
explained in the previous section.

\section{The higher-dimensional case}

The easiest way to study the fuzzy counterpart of diffeomorphism outside of
chiral models is to look at d=1+1 dimensional massive theories. Let us assume
that we know a two-dimensional massive theory in the algebraic formulation
i.e. as an inclusion preserving map of spacetime double cones $\mathcal{O}$
into operator algebras acting on one Hilbert space
\begin{equation}
\mathcal{O}\rightarrow\mathcal{A}(\mathcal{O})
\end{equation}
which fulfill a number of physically motivated requirements among which the
two most prominent ones are Einstein causality and covariance with
energy-momentum positivity. At this point one may as well imagine that this
comes from an algebraic reformulation of the standard formulation in terms of
a collection of pointlike covariant Bose fields which generate the net of
spacetime indexed observable operator algebras. In particular we may want to
look at this net of algebras restricted to the right hand wedge $W:$
$x>\left|  t\right|  .$ Let $\mathcal{A}(W)$ denote the wedge algebra i.e. the
weak limit of the union $\cup_{\mathcal{O}\subset W}\mathcal{A}(\mathcal{O})$
together with its subnet structure. In the setting of classical wave equations
the data inside $W$ are uniquely determined in terms of the characteristic
data along either its upper or lower light ray horizon. The only exception to
this would be the d=1+1 massless theory (classical wave equation) where one
needs the data on both bounding horizons (of the Rindler world $W$), a fact
which harmonizes well with the chiral factorization in the quantum field
theory. With the help of a lightlike translation $T^{(+)}(a)$ along its upper
light ray $R^{(+)}$ we generate a modular inclusion $W_{a}\subset W$ i.e. an
inclusion for which the modular group of $\mathcal{A}(W)$ which is the Lorentz
boost $\Lambda_{t\text{-}x}(\chi)$ for positive rapidities $\chi$ compresses
$\mathcal{A}(W_{a})$ into itself. The relative commutant $\mathcal{A}%
(W_{a})^{\prime}\cap\mathcal{A}(W)$ is an algebra localized on the interval
$(0,a)$ of the right upper lightray; this is the only localization region
which causality associates with this relative commutant. The full upper right
lightray algebra localized on the upper horizon $R^{(+)}$ is obtained as the
union \cite{S-W2}\cite{GLRV}
\begin{align}
\mathcal{A}(R^{(+)})  &  =\bigcup_{t}Ad\Delta_{W}^{i\tau}(A(W_{a})^{\prime
}\cap A(W))\\
\Delta_{W}^{i\tau}  &  =U(\Lambda_{t\text{-}x}(2\pi\tau))\nonumber
\end{align}
with a similar construction of the net $\mathcal{A}(R^{(-)})$ localized on the
lower horizon. It is easy to see that the extension of the halfline net
$\mathcal{A}(R^{(+)})$ to the full line $\mathcal{A}(R)$ (using translations
into the opposite direction or spatial reflections) is a conformal field
theory with almost kinematical aspects \cite{holo} (the spectrum of scale
dimension is integral). The translation and dilation of the 3-parametric
M\"{o}bius group are inherited from the upper light cone translation $T(a)$
and the Lorentz boost $\Lambda_{t\text{-}x}(2\pi\tau)$ and the existence of
the third M\"{o}bius-transformation (the ``rotation'') would follow from the
``standardness of the modular inclusion'' \cite{GLW}. A prerequisite for
obtaining in this way an algebraic version of a chiral QFT is the
nontriviality of the relative commutant from which the nontriviality of the
chiral net would follow. In the classical characteristic propagation setting
for massive modes it is well known that the data on the upper (or lower)
horizon of $W$ determine the wave functions inside $W.$ The counterpart of
this on the quantum level is the equality
\begin{align}
&  \mathcal{A}(W)=\mathcal{A}(R^{(+)})\label{equ}\\
&  \curvearrowright A(R^{1,1})=A(R)\nonumber
\end{align}
(the last line is the equality for the full algebras which follows by using
reflections or opposite translations) which has been proven for massive free
theories in all spacetime dimensions\footnote{For $d>1+1$ the notation
$\mathcal{A}(R^{(+)})$ stands for the transversally unresolved lightfront.
Without the additional transversal resolution the lightfront holography image
is incomplete i.e. one cannot reconstruct the transversal localization
structure of the original net $\mathcal{A}(W)$ even though the equality
(\ref{equ}) continues to hold for the longitudinal part of the net.} by using
the associated Wigner one particle representation theory. For a general proof
we lack sufficient insight into the structure of interacting theories. However
for the class of factorizing d=1+1 theories we are able to show this equality
at least on the level of their formal expansions in terms of bilinear forms
(related to formfactors) \cite{S-W2} \cite{JMP}. From this one learns that the
validity of this equation is not restricted to canonical or near canonical
dimensions of the generating fields although the lightfront fields always must
have (half)integer dimensions \cite{holo}.

The chiral conformal theory $\mathcal{A}(R)$ is not a theory of massless
objects since otherwise it would be impossible to reconstruct the massive
$\mathcal{A}$ theory from its holographic image. A massless chiral theory on
the upper horizon would be invariant against a translation $T^{(-)}(a)$ into
the lower lightray direction but this is definitely not the case for the
present chiral theory (\ref{equ})
\begin{align}
&  \left[  T^{(-)}(a),\mathcal{A}(R^{(+)})\right]  \neq0\\
&  P^{(-)}\cdot P^{(+)}\,\,has\,\,mass\,\,gap\nonumber\\
&  P^{(-)},\,P^{(+)}\,\,are\,\,gapless\nonumber
\end{align}
where the $P^{\prime}s$ are the infinitesimal generators of the two light cone
translations. With other words the chiral theories which originate from
holographic projection onto the horizon possess an additional automorphism
which is not part of the set of physical spacetime automorphisms for zero mass
theories. The prerequisite for an upper/lower chiral conformal theory namely
the gaplessness of lightcone translations is always guarantied. In order to
recover the full $\mathcal{A}(W)$ theory in higher spacetime dimensions
$d>1+1$ the study of only one modular inclusion is not sufficient. We will
present the necessary steps for an invertible holographic imaging somewhere
else (for some ideas in this direction see ).

Before we use the above d=1+1 holographic projection for the construction of
higher modular symmetries it is instructive to contrast the above construction
of a horizon-localized theory with the formal ``lightcone quantization'' or
``infinite momentum frame'' method. In that case one selects concrete field
coordinates and studies what happens to them of one restricts to the lightray
without converting them first into the net of algebras which they generate.
One runs into problems even if these fields are free fields which are related
to the singularities which one encounters if one approaches a light ray
distance in pointlike correlation functions of pointlike fields. Whereas some
of these problems (at least in the case of low spin) can be avoided by careful
application of distribution theory, this is not possible if the ultraviolet
singularities are too big. The reprocessing of a field into its associated net
of field-coordinatization-free algebras is a much more radical physical step
than enhancing mathematical rigor by the use of distribution theory. Our above
arguments have shown that the nontriviality of relative commutant algebras in
the modular inclusion construction has a much better chance than nontriviality
in the face of hard to control ultraviolet behavior. After the holographic net
has been constructed one may parametrize this net in terms of pointlike chiral
fields; the possibility of generating chiral nets by pointlike fields is
guarantied by mathematics \cite{Joerss} and again there is no limitation from
short distance behavior. But these chiral fields after holographic projection
have little to do with the original fields (assuming that our original net of
algebras was generated by massive pointlike fields). This shows why lightcone
quantization remained a highly artistic (mathematically hard to control, but
with a lot of physical intuitive content) procedure. What is behind it is
``holographic reprocessing'' linking QFTs in different spacetime dimensions,
whose understanding defies the methods of standard QFT as
Lagrangian/functional integral quantization but not their physical principles.

Having arrived at a faithful (isomorphic) holographic image, it is not
difficult to guess what the looked for fuzzy analogs of the chiral
diffeomorphisms are. Certainly the conformal rotation acts on the
$\mathcal{A}(R)$ net geometrically whereas its action on the two-dimensional
massive net $\mathcal{A}(R^{1,1})$ is fuzzy. But according to the previous
section there is the infinite group of chiral diffeomorphisms which.

As a result of the necessity to obtain as well the transversal net structure,
the higher dimensional holography is much more involved and will not be
discussed here (some details can be found in \cite{holo}).

\section{Outlook}

In this short note we tried to draw attention to a new point of view which may
be important for a better understanding of QFT as well as for a possible
mathematical enrichment in the area of infinite dimensional Lie groups. By
using appropriately defined holographicc relations with chiral algebras on the
lightfront horizons we argued that this is the correct analog of the
diffeomorphisms in chiral theories.

The important mathematical tool in all these considerations is the modular
theory of Tomita-Takesaki adapted to the field theoretic nets of local von
Neumann algebras. These are the same methods \cite{JMP} as those which show
the thermal aspects of the vacuum state upon restriction to local algebras of
which the Unruh-Hawking illustration for wedge-localized algebras is the most
prominent example. Our findings do not only generalize the spacetime Noether
symmetries of the Lagrangian formalism, but they also make the recent
(somewhat ad hoc) use of the chiral diffeomorphism algebras in connection with
attempts at the derivation of a sufficiently generic (i.e. model independent)
quantum Bekenstein area law more palatable\textit{\cite{Carlip}. }In fact they
suggest to look for a ``localization entropy'' in standard QFT which could be
the Minkowski space local quantum physics analog which preempts the existence
of a sufficiently generic quantum explanation of the Bekenstein area law in
the more geometric setting of curved spacetime and eventually may lead to
clues about quantum gravity. The almost kinematical nature of these
holographic lightfront algebras \cite{holo} makes the universal aspect of an
area law plausible.

Whereas the thermal aspect of wedge- or double- cone localized algebras has a
very easy explanation\textit{\ }in terms of the AQFT setting, the
localization- entropy discussion is much more subtle \cite{JMP}. By the very
nature of the local von Neumann algebras (hyperfinite type III$_{1}$) their
(von Neumann) entropy with respect to any state is always infinite. One needs
the formalism of ``split inclusion'' in order to arrive at the conjecture that
perhaps the restriction of the vacuum state to the local algebra including a
``collar'' of thickness $\delta$ around the boundary may be finite and for all
sufficiently small $\delta$ proportional to the area in a way which diverges
for $\delta\rightarrow0.$ In this case the mechanism for the localization
entropy would be similar to the vacuum polarization effect in partial Noether
charges with its proportionality to the area and its divergence for
testfunctions whose collar size (the region where the smearing function
decreases from its value one to zero) shrinks to zero.

>From such a localization entropy with area proportionality towards a finite
geometric entropy in the classical sense of Bekenstein and Hawking there may
be still a long way. But what seems to be very intriguing already now is the
observation that attempts of AQFT at generalizations of symmetries, encoding
of spacetime into algebraic relations, thermal and entropy aspects of
localization, holography and transplantations of QFTs point use the same
modular concepts as the ones in the present work.

\textit{Acknowledgement}: one of the authors (B. S.) thanks Jakob Yngvason for
an interesting and helpful correspondence.

\textit{Note added}: In a recent preprint \cite{Fard} the generalization of
the state vector $\Phi$ to the higher analogs $D_{n},$ $n>2$ of the dilation
were computed. These results are correct, if suitably restricted to one of the
multi-localized algebras as done in the present paper (even though the
underlying assumption of faithfulness of these states on the multi-local
algebras is incorrect). There will be a revised version of \cite{Fard} in
which the split property is used in order to obtain faithful and at the same
time geometrically useful states on multi-localized algebras.

\subsection{Appendix: Examples from d=1+1 holography}

Before presenting some examples it may be helpful to remind the reader that
although in d=1+1 massive theories lightray restrictions as well as zero mass
limits of fields lead both to chiral theories one should avoid confusing these
limits. In order to avoid infrared divergences consider a (selfconjugate) free
field with a halfinteger ``Lorentz-spin''
\begin{align}
\psi(x)  &  =\frac{1}{\sqrt{2\pi}}\int\left(  e^{ipx}u(\theta)a(\theta
)+e^{-ipx}v(\theta)b^{\ast}(\theta)\right)  d\theta=\left(
\begin{array}
[c]{c}%
\psi_{+}\\
\psi_{-}%
\end{array}
\right) \\
u(\theta)  &  =\sqrt{m}\left(
\begin{array}
[c]{c}%
e^{-\frac{1}{2}\theta}\\
e^{\frac{1}{2}\theta}%
\end{array}
\right)  ,\,\,v(\theta)=\sqrt{m}\left(
\begin{array}
[c]{c}%
e^{-\frac{1}{2}\theta}\\
-e^{\frac{1}{2}\theta}%
\end{array}
\right) \nonumber\\
p  &  =m(ch\theta,sh\theta)\nonumber
\end{align}
The zero mass limit for the first component requires
\begin{align}
&  lim_{m\rightarrow0}e^{\frac{1}{2}(lnm-\theta)}<\infty\\
&  \curvearrowright\theta=\hat{\theta}+lnm\rightarrow\hat{\theta}%
-\infty\nonumber
\end{align}
i.e. $\theta$ is driven towards (negative) infinity. The c-number factors in
the rapidity integral can be rewritten in terms of the reduces rapidity
$\hat{\theta}$ only
\begin{equation}
e^{ipx}\sqrt{m}e^{-\frac{1}{2}\theta}\rightarrow e^{ie^{\hat{\theta}}x_{+}%
}e^{^{-\frac{1}{2}\hat{\theta}}}%
\end{equation}
but there is no limit for the annihilation/creation operators in the Fock
space $a(\theta)\rightarrow a(\hat{\theta}-\infty).$ Although there is no zero
mass limit in the original Fock space, such a limit makes perfect sense for
the two-point function (and hence for all correlation functions). Its operator
reconstruction gives a radical change of Fock space: instead of the original
$a^{\#}-b^{\#}$ degrees of freedom one now has the well-known
\textit{left-right mover doubling}.

The lightray restriction on the other and leads to a chiral theory without
this doubling. Parametrizing the x in terms of \ Lorentz-length $\sqrt{\pm
x^{2}}$ and rapidity $\chi,$ and letting the $ln\sqrt{\pm x^{2}}$ approach
$-\infty$ and $\chi\rightarrow\infty$ (which leaves the argument of the
operators unchanged), we obtain a chiral field without doubling. The
2-dimensional massive field may be reproduces by applying a lightray
translation in the opposite direction.

We also may consider the last limit in the spirit of the more rigorous
holographic setting. In that case one would start from the definition of the
wedge localized algebra which can be generated by smeared fields with smearing
functions supported in $W$%
\begin{equation}
\psi_{+}(f),\,\,suppf\in W \label{alg}%
\end{equation}
and fulfill a modular boundary condition in their strip-analytic
(rapidity-parametrized) Fouriertransforms which characterize the real subspace
$H_{R}$ of the previous section. The modular theory for Fermions is the same
up to an additional ``twist'' operator $K$ in $S=JK\Delta^{\frac{1}{2}}$ which
accounts for the difference between the geometric and the algebraic opposite
(in the sense of the commutant). The same algebra can be generated with the
second component $\psi_{-}(g);$ one only has to pay attention to the slightly
different analytic characterization of $H_{R}$ in terms of the $g^{\prime}s.$
In this case the upper horizon algebra defined as in the previous section can
also be generated by smearing the chiral lightray fields obtained in the limit
$ln\sqrt{-x^{2}}\rightarrow-\infty,\,\chi\rightarrow\hat{\chi}+\infty$ with
test functions $f(x_{+})$ $suppf\in R_{+}$ whose rapidity-parametrized
Fouriertransforms with the modular boundary conditions define the same spaces
as those in (\ref{alg}). With other words not only the algebras (\ref{equ})
are identical, but already the generators are equal. It is very important to
stress that the direct construction of such lightray-restricted fields only
works if the operator dimension of the field is close to its free-field value.
More precisely the two-point Kallen-Lehmann spectral function must fulfill
\begin{equation}
\int\rho(\kappa^{2})d\kappa^{2}<\infty
\end{equation}
which is the as the well known finiteness of the inverse wave function
renormalization constant of the field in the canonical formalism. In addition
to this restriction there is an infrared restriction on the smearing function
for small momenta in the case of s=0
\[
\int_{0}^{\infty}\frac{\bar{f}(p)g(p)}{p}dp<\infty
\]
which restricts the localizing test functions in x space to those which are a
derivative of a localized function (so that only the current i.e. the
derivative of the would be scalar field makes sense). In free field theory the
above restriction process is therefore only applicable to the free field
itself, but not to its composites which constitute the Borchers equivalence
class of field-coordinatizations.

In the presence of interactions one cannot expect pointlike fields which
generate the chiral holographic projection to be simply related to the
original field. It is known that the Zamolodchikov-Faddeev nonlocal
generalization of the free field algebra (we only look at the simplest case
for which the so-called Sinh-Gordon Z-F algebra is an example)
\begin{align}
F(x)  &  =\int\left(  Z(\theta)e^{-ipx}+Z^{\ast}(\theta)e^{ipx}\right)
d\theta\\
Z(\theta)Z(\theta^{\prime})  &  =S(\theta-\theta^{\prime})Z(\theta^{\prime
})Z(\theta)\nonumber\\
Z(\theta)Z^{\ast}(\theta^{\prime})  &  =S^{-1}(\theta-\theta^{\prime})Z^{\ast
}(\theta^{\prime})Z(\theta)+\delta(\theta-\theta^{\prime})\nonumber
\end{align}
where the $Z^{\#}s$ can be explicitly represented in terms of nonlocal
rapidity space expressions in $a^{\#\prime}s$ Fock space operators. It has
been shown that although the $F(x)$ are non-causal (non-local), they still
carry the notion of wedge-localization i.e. $F(f)$ is affiliated with the
wedge algebra $\mathcal{A}(W)$ if and only if the structure functions of the
Z-algebra fulfill the properties of an admissable 2-particle elastic S-matrix
i.e. besides unitarity the all important crossing symmetry including a
bootstrap bound state structure. In fact it is believed that the spacetime
concept of d=1+1 tempered polarization-free generators \cite{BBS} (PFG's) and
the momentum space Z-F algebraic structure are completely equivalent. A
general $\mathcal{A}(W)$-affiliated operator can be shown to be of the form%

\begin{equation}
A=\sum\frac{1}{n!}\int_{C}...\int_{C}a_{n}(\theta_{1},...\theta_{n}%
):Z(\theta_{1})...Z(\theta_{n}): \label{series}%
\end{equation}
where the $a_{n}$ have a simple relation to the various formfactors of $A$
(including bound states) whose different in-out distributions of momenta
correspond to the different contributions to the integral from the upper/lower
rim of the analytic strip bounded by the contour $C$, which are related by
crossing \cite{JMP}. Note that when we write down such sums without discussing
their convergence related to their operator aspects we are only dealing with
bilinear forms i.e. matrix elements between a dense set of multi-particle
state vectors of ``would-be operators''. The affiliation to double cones (Fig
2) is obtained by restricting the space of wedge affiliated objects by
requiring their commutance with the spatially translated (to the right inside
$W$) generators $F_{a}(f)$%
\begin{align}
\left[  A,F_{a}(f)\right]   &  =0\,\,\forall f\text{\thinspace\thinspace
}suppf\subset W\\
F_{a}(f)  &  =U(a)F(f)U^{\ast}(a)\nonumber
\end{align}
This condition defines the bilinear forms $A$ affiliated with $\mathcal{A}%
(C_{a})=\mathcal{A}(W_{a})^{\prime}\cap\mathcal{A}(W)$ which for $a=$right
spacelike is a double cone $C_{a}$ with one apex in the origin and the other
at $a.$ The condition on the coefficients is the famous ``kinematical pole
relation'' of Smirnov \cite{Smir}, a recursive condition which reads%

\begin{equation}
a_{n+2}(\vartheta+i\pi+i\varepsilon,\vartheta,\theta_{1},\theta_{2}%
....\theta_{n})\simeq\frac{1}{\varepsilon}\left[  1-\prod_{i=1}^{n}%
S(\vartheta-\theta_{i})\right]  a_{n}(\theta_{1},\theta_{2}....\theta_{n})
\label{pole}%
\end{equation}
with an additional Paley-Wiener type condition on the meromorphic $a_{n}s$
which encodes the geometry of $C_{a}.$ Pointlike fields obey the same relation
but with a polynomial behavior which distinguishes the individual fields after
having split off certain standard formfactors which are the same for all
fields in the same sector.

The important point in all these technicalities is now the fact that if we
would have taken a lightlike $a$ on the upper horizon of $W$ as was done for
the definition of an interval algebra (Fig 1)
\[
A(R(0,a))=\mathcal{A}(W_{a})^{\prime}\cap\mathcal{A}(W)
\]
in the holographic formalism of section 4, the condition would also had led us
to (\ref{pole}). This underlines the equality (\ref{equ}) of the wedge algebra
with its holographic image . But the would be pointlike fields affiliated with
$\mathcal{A}(W)$ have no restrictions onto the upper horizon, rather the
chiral fields affiliated with $\mathcal{A}(R_{+})$ have to be determined anew
according to \cite{Joerss} using the M\"{o}bius symmetry of the holographic
image. As always suspected the chiral theories on the lightray are very
simple; as a result of their (half)integer spectrum of scale dimensions they
have no genuine dynamical properties and the dynamical richness of the
original theory lies in the rich action of symmetries (which are absent in
chiral theories obtaines as the nassless limit of massive ones) which are
needed in order to return to the ambient theory. This is made possible by the
fact that the spacetime indexing of the holographic net (and hence the
physical interpretation) is very different (relatively fuzzy) to that of the
original net. Realistic lightfront holography implies a violent
``scrambling'', but it is still possible to reconstitute the original net
structure. In this respect the very mild AdS-CQFT holography is somewhat atypical.

\end{document}